\documentclass[conference, a4paper]{IEEEtran}
\IEEEoverridecommandlockouts

\usepackage{textcomp}
\makeatletter
\def\endtable{\end@float}
\def\endfigure{\end@float}
\makeatother
\usepackage{tikz}
\usepackage{tabularx}
\usepackage{cite}
\usepackage{array}
\usepackage{graphicx}
\usepackage[cmex10]{amsmath}
\usepackage{mathtools}
\usepackage{multirow}
\usepackage{gensymb}
\usepackage{subfigure}
\usepackage{nomencl}
\usepackage{color}
\usepackage{float}
\usepackage{balance}
\usepackage{amssymb}
\usepackage{hyperref}
\usepackage{colortbl}
\usepackage{tabularray}
\usepackage{ctable}
\usepackage{notoccite}
\usepackage{acronym}

\usepackage{comment}

\usepackage{flushend}
\usepackage{algorithm}
\usepackage[noend]{algpseudocode}

\newcommand{\mr}{\mathrm}

\usepackage{dcolumn}
\newcolumntype{d}[1]{D{.}{.}{#1}}

\newcolumntype{L}[1]{>{\raggedright\let\newline\\\arraybackslash\hspace{0pt}}m{#1}}
\newcolumntype{C}[1]{>{\centering\let\newline\\\arraybackslash\hspace{0pt}}m{#1}}
\newcolumntype{R}[1]{>{\raggedleft\let\newline\\\arraybackslash\hspace{0pt}}m{#1}}

\acrodef{der}[DER]{distributed energy resource}
\acrodef{res}[RES]{renewable energy source}
\acrodef{sg}[SG]{synchronous generator}
\acrodef{ibr}[IBR]{inverter-based resource}
\acrodef{pll}[PLL]{phase-locked loop}
\acrodef{avr}[AVR]{automatic voltage regulator}
\acrodef{pcc}[PCC]{point of common coupling}
\acrodef{svm}[SVM]{support vector machine}
\acrodef{pi}[PI]{proportional-integral}
\acrodef{asm}[ASM]{adaptive sampling method}
\acrodef{scr}[SCR]{short-circuit ratio}

\graphicspath{{Figs/}}
\makeatletter
\newcommand{\linebreakand}{%
  \end{@IEEEauthorhalign}
  \hfill\mbox{}\par
  \mbox{}\hfill\begin{@IEEEauthorhalign}
}

\newcommand\copyrighttext{%
  \footnotesize
  \centering Paper submitted to 9th International Conference on CLEAN ELECTRICAL POWER, ICCEP 2025.}
\newcommand\copyrightnotice{%
\begin{tikzpicture}[remember picture,overlay]
\node[anchor=south,yshift=0pt] at (current page.south) {\setlength{\fboxrule}{0pt}\fbox{\parbox{\dimexpr\textwidth-\fboxsep-\fboxrule\relax}{\copyrighttext}}};
\end{tikzpicture}%
}

\begin{document}

\title{Detailed Small--Signal Stability Analysis of the \\ 
Cigr\'{e} High--Voltage Network Penetrated by\\
Grid--Following Inverter--Based Resources \thanks{This work has been funded by the EU fund Next Generation EU, Missione 4, Componente 1, CUP D53D23001650006, MUR PRIN project 2022ZJPPSN SCooPS.}}

\author
{

\IEEEauthorblockN{Francesco Conte\IEEEauthorrefmark{1},
Fernando Mancilla-David \IEEEauthorrefmark{2},
Amritansh Sagar \IEEEauthorrefmark{1}, \\
Chendan Li \IEEEauthorrefmark{3}, Federico Silvestro \IEEEauthorrefmark{3}, and Samuele Grillo \IEEEauthorrefmark{2}
}
\\
\IEEEauthorblockA{\IEEEauthorrefmark{1} Facoltà Dipartimentale di Ingegneria, Università Campus Bio-Medico di Roma, Rome, Italy.\\ \{f.conte, a.sagar\}@unicampus.it)}
\IEEEauthorblockA{\IEEEauthorrefmark{2} Dipartimento di Elettronica, Informazione e Bioingegneria, Politecnico di Milano, Milan, Italy.\\
\{fernandoadolfo.mancilla, samuele.grillo\}@polimi.it)}
\IEEEauthorblockA{\IEEEauthorrefmark{3} Dipartimento di Ingegneria Navale, Elettrica, Elettronica e delle Telecomunicazioni,\\
Università degli Studi di Genova, Genoa, Italy.\\
\{federico.silvestro, chendan.li\}@unige.it)}
}

\IEEEaftertitletext{\copyrightnotice\vspace{0.2\baselineskip}}

\maketitle
\begin{abstract}
This paper presents a detailed small-signal stability analysis of a modified version of the Cigr\'{e} European high-voltage network, where one of the synchronous generators is replaced by a grid-following inverter-based resource (IBR). The analysis focuses on the influence of the parameters defining the grid-following IBR control scheme on the stability of the system. Given a set of potential grid configurations and the value of the IBR control parameters, stability is verified by the direct eigenvalue analysis of a high-detailed linearized model of the overall Cigr\'{e} network. Starting from this procedure, we propose an adaptive sampling method for training a support vector machine classifier able to estimate the probability of stability of the power system over a domain defined by candidate intervals of the considered parameters. The training of the classifier is refined to identify with more accuracy the boundaries of the parameters' stability regions. The obtained results are then compared with those obtained by representing the grid with the classical Thévenin equivalent. Results suggest that, when the Thévenin equivalent is accurate, the predicted stability region is conservative yet contained within that of the full network.    
\end{abstract}


\IEEEpeerreviewmaketitle


\section{Introduction}
\label{sec:intro}

The transition to the extensive use of \acp{res} is recognized as a key challenge  of the current century. The effectiveness of this transformation significantly depends on the accessibility of methodologies and procedures that facilitate the affordable, resilient, and technically sustainable integration of \ac{res} in the power system. One of the key difference between traditional fossil-fueled generators and \ac{res} is that the former are \acp{sg}, whereas the latter are \acp{ibr}, i.e., power sources connected to the grid by a power converter. Unlike \acp{sg}, \acp{ibr} lack spinning masses that provide natural inertia to the system and rely on specific control mechanisms to synchronize with the grid and regulate the power exchange. This generally makes the response of \acp{ibr} to system perturbations significantly different from that of a \ac{sg} and extremely dependent on the converter control design. 

Focusing on power system stability, this means that the progressive replacement of \acp{sg} with \ac{ibr} constitutes a momentum change, leading to a revision of the definition of power system stability, as carried out in \cite{Hatziargyriou2021}. In this work, authors propose an extension of the power system stability analysis by adding to the classical rotor angle, voltage and frequency stability the “resonance stability” and the “converter-driven stability”.

In particular, converter-driven stability concerns the interactions of the control systems of \acp{ibr} with other power system components such as: interaction of the inner-current loops of \acp{ibr} with passive power system components; interaction of
the converter filter with converter high-frequency switching; interaction among nearby converter controllers; interaction of the electromechanical dynamics of \ac{sg} with \acp{ibr} outer control loops and \acs{pll}. The consensus is that these new dynamical interactions arising in the grid have an impact on electromechanical stability even if they may lead to instabilities in the timescale from tens of nanoseconds to tens of milliseconds, neglected in the classical rotor angle, frequency and voltage stability analyses. 

Most of the existing literature study the stability of different classes of \acp{ibr} by modeling the network with an infinite bus (an ideal voltage source) connected to the converter through the overall grid impedance. This well known simplified representation of the network is also known as Th\'{e}venin equivalent. For example, \cite{Rodriguez2019,Collados2020} study the small-signal stability of grid-following \acp{ibr} connected to a Th\'{e}venin equivalent also considering the impact of dead-time and time delays. The authors of \cite{Huang2022} and \cite{Zhang2024} focus on the impact of the \ac{pll} dynamics in the small-signal stability of grid-following \ac{ibr} connected to ``weak'' grids (where weakness is usually represented by a high grid impedance). Other examples of small-signal stability analyses always by modeling the network with a Th\'{e}venin equivalent considering grid-following \acp{ibr} can be found in \cite{Ma2024,Saleem2023,Saleem2024}. The Th\'{e}venin equivalent is also adopted in further papers, such as \cite{Mohammed2024}, to analyze the stability of grid-forming \acp{ibr}. 
  
Such a widespread use of the Th\'{e}venin equivalent has certainly enabled the discovery of numerous relationships between inverter stability and its control parameters, as a function of different grid conditions, usually represented by varying values of the grid impedance. Although highly relevant, this approach does not allow for the evaluation of the aspects previously mentioned, such as the interaction between different \ac{ibr} connected to the grid or between \acp{ibr} and \acp{sg}.

For this reason, some recent works, such as \cite{Collados2022} and \cite{Ding2025} propose small-signal stability analyses of power systems with \acp{ibr} by using more comprehensive and accurate models of the grid. One of the main issues with such approaches is their computational complexity, as small-signal analysis requires linearizing the entire system and computing the corresponding eigenvalues. For example, studying the relationship between a control parameter and system stability may require a huge repetition of this procedure.

In the present paper, we present the preliminary results of a methodology developed to analyze the stability of power systems with high penetration of \acp{ibr} adopting a high-detailed modeling of the overall system. The basic procedure to assess stability is the direct eigenvalue analysis of the linearized full grid model. This model can be computed for a given grid configuration--which, in a real-world scenario, may vary if any subunit of the \acp{sg} is out of service and/or depending on the operating points of loads and generators—-and for a specific set of parameters under investigation. Focusing on the control parameters of the \acp{ibr}, we propose an \ac{asm} for training a \ac{svm} probabilistic classifier able to estimate the probability of stability of the power system as a function of a set of parameters. In this way, the computational complexity required to study the impact of the control parameters on stability is significantly reduced. The training of the classifier is refined to identify with more accuracy the boundaries of the parameters' stability regions, allowing to define proper stability conditions.  

The proposed method is applied to a modified version of the Cigr\'{e} European HV transmission network introduced in \cite{cigre}, where one of the \ac{sg} is replaced by a grid-following \ac{ibr}. The obtained results are compared with those obtained by representing the grid with the Th\'{e}venin equivalent. 

The rest of the paper is organized as follows. Section~\ref{sec:Cigre_network} introduces the modified Cigr\'{e} network; Section~\ref{sec:stability_analysis_method} describes the stability analysis method; Section~\ref{sec:results} provides and discusses the obtained results; finally, Section~\ref{sec:Conclusions} reports the concluding remarks.

\section{Modified Cigr\'{e} network}
\label{sec:Cigre_network}
\begin{figure}[!t]
    \centering
    \includegraphics[width=1\columnwidth]{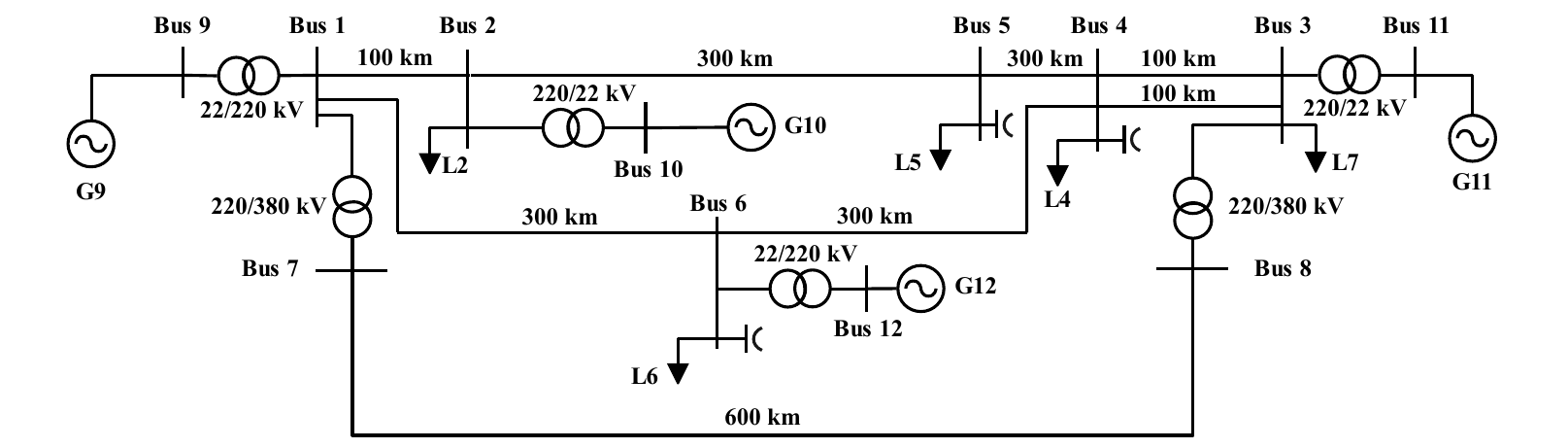}
    \caption{Modified Cigr\'{e} European HV network schematic.}
    \label{fig:cigre}
\end{figure}

The case study considered in this paper is the benchmark Cigr\'{e} European HV transmission network introduced in \cite{cigre}. The system topology and composition are shown in Fig.~\ref{fig:cigre}. Transmission voltages are 220~\texttt{kV} and 380~\texttt{kV}; the system frequency is 50~\texttt{Hz}. The network has 12 buses; four \acp{sg} are connected at the 22~\texttt{kV} buses 9,10,11 and 12. The adopted parameters basically are the ones indicated in \cite{cigre} with the following modifications and integrations.

Parameters changes:
\begin{itemize}
    \item the resistance of the transformers is set to 2\% (in the base case it is zero);
    \item the nominal power of the 22/220 \texttt{kV} transformers allowing the connection of generators is set equal to the one of the relevant generator (in the base case, the transformers nominal powers are higher);
    \item the nominal power of generator G10 is set to 350~\texttt{MVA} (in the base case it is equal to 700~\texttt{MVA}).
\end{itemize}

Integrations:
\begin{itemize}
    \item the generator G9 is modeled as a generator with the same per unit parameters of G11 and with a nominal power equal to 620~\texttt{MVA} (in the base case it is modeled as an ideal voltage source);
    \item all \ac{sg} are equipped with the steam turbine and governor model \cite{mathworks2024steam,governor}, operating 5\% frequency droop regulation, and with the excitation system with the IEEE type 1 \ac{avr} as modeled in \cite{mathworks2024excitation,avrIEEE} (the Cigr\'{e} report \cite{cigre} does not provide any indication about governors or excitation systems).
\end{itemize}

The main objective of this paper is to study the stability of this system assuming that generator G10 is replaced by an equivalent \ac{ibr} with the same nominal power. The details of the considered \ac{ibr} are provided in Section~\ref{ssec:cig}. 

In order to study the stability in different network configurations, starting from the established base case, we designed the four scenarios presented in Table~\ref{tab:scenarios}, where $S^{nom}_i$, $i=9,10,11,12$ indicate the SGs nominal power, $S_G^{nom}$ indicates the total nominal power of the grid generators, and $S_L^{peak}$ is the peak load. The latter is defined according to the indications provided in the Cigr\'{e} report \cite{cigre}. It is worth remarking that scenario~1 is the base case, where all generators have their full nominal power; in the other scenarios, the size of one of the generators, excepting to G10 (to be replaced by the equivalent \ac{ibr}), is divided by two. The assumption is that the \acp{sg} are actually composed of two subunits and thus scenarios~2-4 considers the case where one of them is out of service, for example, due to maintenance.    
\begin{table}[t]
\caption{Generators nominal power (in \texttt{MVA}) in the various network configuration scenarios.}
\label{tab:scenarios}
\begin{center}
\begin{tabular}{ccccccc}
\hline \\
\vspace{-15pt} \\
Scenario & $S_9^{nom}$ & $S_{10}^{nom}$ & $S_{11}^{nom}$ & $S_{12}^{nom}$ & $S_G^{nom}$ & $S_L^{nom}$ \\
\vspace{-7pt} \\
\hline
\vspace{-5pt} \\
1 & 620 & 350 & 500 & 500 & 1970 & 1600 \\
2 & 310 & 350 & 500 & 500 & 1660 & 1600 \\
3 & 620 & 350 & 250 & 500 & 1720 & 1600 \\
4 & 620 & 350 & 500 & 250 & 1720 & 1600 \\ \hline
\end{tabular}
\end{center}
\end{table}

It is worth noting that in all scenarios the load is maintained at its peak value, and that the base case generator sizes have been determined to ensure feasible operating points, which are provided in Table~\ref{tab:dispatches}.

\begin{table}[t]
\caption{Generators operating points in the various network configuration scenarios.}
\label{tab:dispatches}
\begin{center}
\begin{tabular}{ccccc}
\hline \\
\vspace{-15pt} \\
Generator & $G9$ & $G10$ & $G11$ & $G12$ \\
\vspace{-7pt} \\
\hline
\vspace{-5pt} \\
$V_0$  \texttt{pu} & 1.02 & 1.02 & 1.03 & 1.02 \\
\vspace{-7pt} \\
\hline
\vspace{-5pt} \\
\multicolumn{5}{c}{Active power set points $P_0$ \texttt{MW}}  \\
\vspace{-7pt} \\
\hline
\vspace{-5pt} \\
Scenario 1  & 503 & 241 & 345 & 345  \\
Scenario 2 & 222 & 316 & 451 & 451  \\
Scneario 3 & 530 & 133 & 176 & 397 \\
Scenario 4 & 546 & 278 & 397 & 176  \\ \hline
\end{tabular}
\end{center}
\end{table}

\subsection{IBR model}
\label{ssec:cig}
\begin{figure}[!t]
\centering
\includegraphics[width=1\columnwidth]{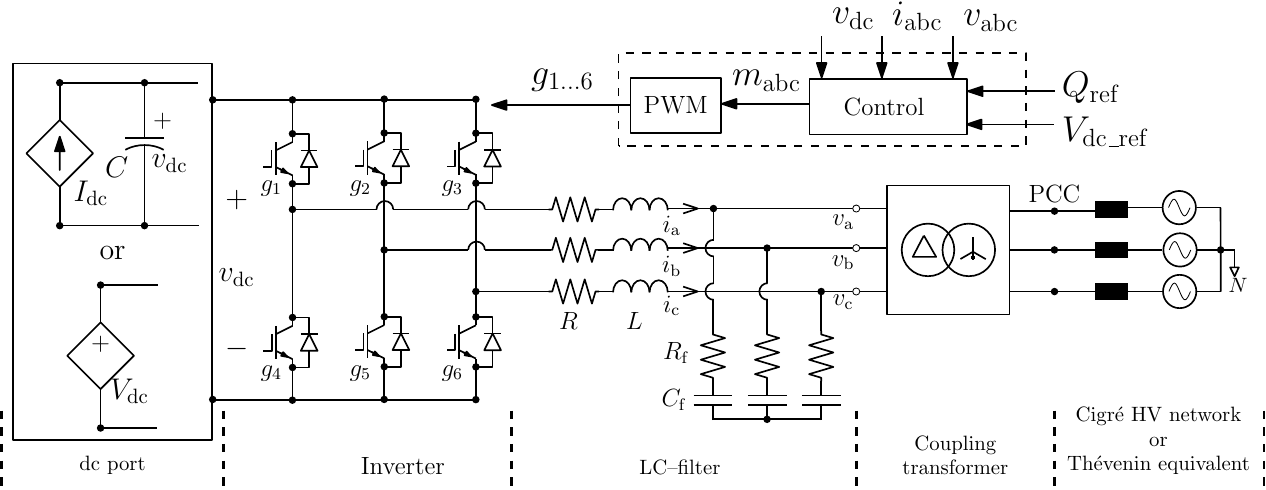}
\caption{Schematic of the two-level three-phase \ac{ibr} considered in the paper.}
\label{fig:VSI_power}
\end{figure}

\begin{figure}[!t]
\centering
\includegraphics[width=1\columnwidth]{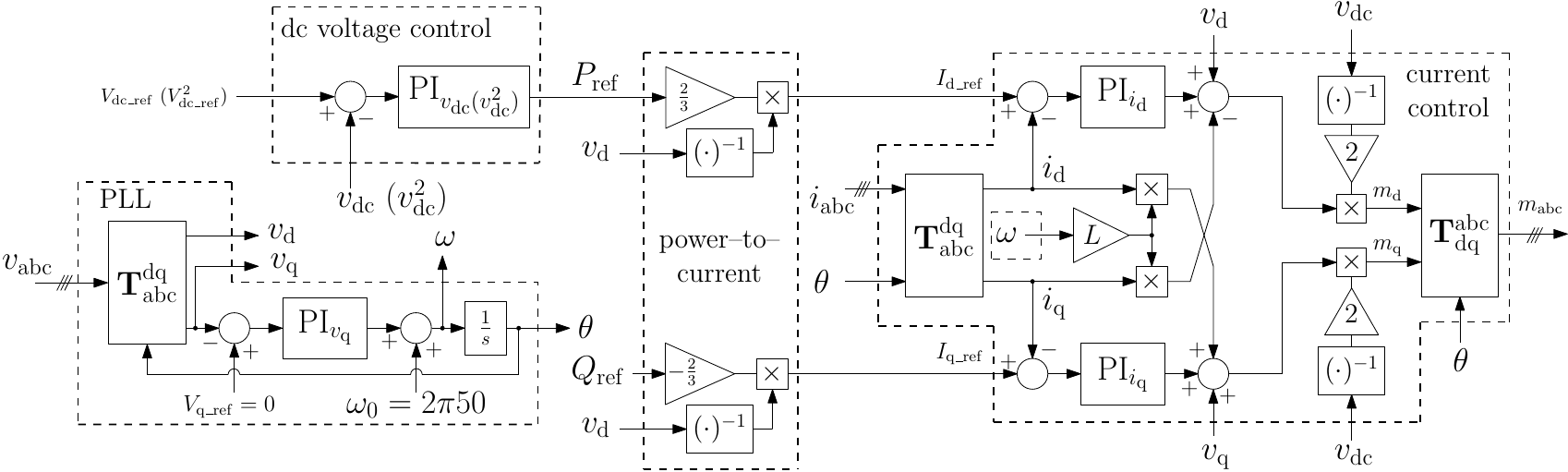}
\caption{Block diagram of the IBR control scheme.}
\label{fig:VSI_control}
\end{figure}

The modeling approach for \acp{ibr} is based on current market availability. To the best of the authors' knowledge, the largest commercially available inverters for \ac {ibr} applications are those manufactured by GamesaElectric. These are rated at 5~\texttt{MVA}, 1500~\texttt{V} on the dc side, and synthesize 660~\texttt{V} / 50~~\texttt{Hz} on the ac side. They are based on a two-level, three-phase voltage source converter architecture \cite{gamesa}. It is further assumed that they use pulse width modulation at 6~\texttt{kHz}, and feature a capacitor on the dc port and an LC filter on the ac port. Fig.~\ref{fig:VSI_power} illustrates the full realization, including a standard 5~\texttt{MVA} $\Delta$-$Y_n$ coupling transformer. The control scheme, illustrated in Fig.~\ref{fig:VSI_control}, consists of four distinct control stages. 

\begin{figure}[!ht]
    \centering
    \includegraphics[width=0.7\linewidth]{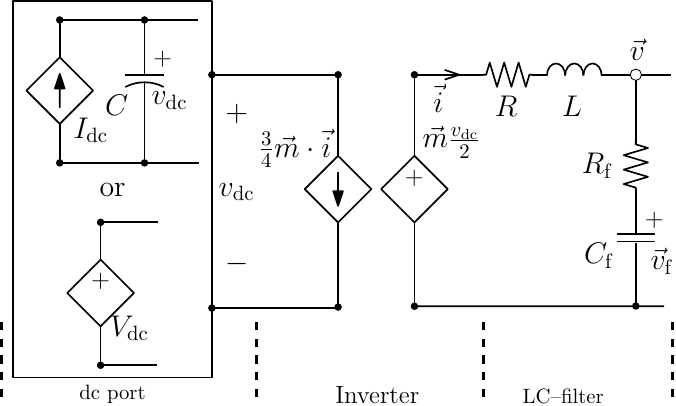}
    \caption{\ac{ibr} equivalent circuit schematic in the $dq$ domain.}
    \label{fig:controlsources}
\end{figure}

The current control, implemented in a synchronous $dq$ frame, decouples the control of the current on each $d$- and $q$-axis by means of feedforward loops, driving them to reference values using identical \ac{pi} regulators, and translates the current references into active and reactive power references via a power-to-current transformation \cite{yazdani2010}. This approach relies on a stiff voltage at the \ac{pcc} and on forcing the $q$-axis voltage to be zero. The latter is achieved by the \ac{pll}, which synchronizes to the grid by measuring the voltage at \ac{pcc}, transforming it into the $dq$ frame, and zeroing out $v_q$ via a PI controller \cite{chung2000phase}. Finally, the control scheme includes regulation of the dc-port voltage to a desired reference. This can be done by directly controlling the voltage across the $dc$ capacitor, i.e., by controlling $v_\mr{dc}$, or by controlling the energy stored in the capacitor, i.e., by controlling $v_\mr{dc}^2$ \cite{yazdani2010}. In both cases, the control action is performed by a \ac{pi} regulator. 

In summary, the control scheme relies on selecting appropriate gains for three \ac{pi} controllers: those for current control, the \ac{pll}, and the dc-port voltage control.

A 350~\texttt{MVA} power plant is modeled by aggregating 70 of the aforementioned 5~\texttt{MVA} units, following the procedure introduced in \cite{Mishra2024} and assuming identical behavior on the dc port. 

\section{Small-signal stability analysis method}
\label{sec:stability_analysis_method}

\subsection{Model linearization}
\label{ssec:linearization}
In this paper, power system stability analysis is carried out through the direct eigenvalues analysis of the full linearized model. In the following, we briefly introduce the adopted linearization method.

The complete linearization is realized in the dq reference frame. First, each subsystem modeling a network component (\acp{sg}, \acp{ibr}, loads,...) is linearized and expressed in the state-space form after having properly defined input and output vectors. Then, the full model is obtained by connecting the linear subsystems. This method is the same proposed in \cite{Collados2020,Collados2024}.

Table~\ref{tab:input-output-linearization} summarizes the inputs and outputs defined for each component. Here, $\vec{v}$ and $\vec{i}$ are the voltage and current phasors at the point of connection with the other network components (e.g., the \ac{pcc} for \acp{sg} and \ac{ibr}). Both $\vec{v}$ and $\vec{i}$ are defined with respect to the same global reference frame (the one of Bus 9 in the Cigr\'{e} network).

\begin{table}[t]
\caption{Input-output definition of the network components subsystems.}
\label{tab:input-output-linearization}
\begin{center}
\begin{tabular}{lcc}
\hline
\textbf{Component} & \textbf{Inputs}                                & \textbf{Outputs} \\ \hline
\ac{sg}            & $P_{\rm ref},V_{\rm ref},\vec{i}$              & $\vec{v}$        \\
\ac{ibr}           & $P_{\rm ref}(I_{\rm dc}), Q_{\rm ref},\vec{i}$ & $\vec{v}$        \\
Transformer        & $\vec{v}$                                      & $\vec{i}$        \\
RL-branch          & $\vec{v}$                                      & $\vec{i}$        \\
RL-load            & $\vec{v}$                                      & $\vec{i}$        \\
Shunt capacitor    & $\vec{i}$                                      & $\vec{v}$        \\ \hline
\end{tabular}
\end{center}
\end{table}

Some details of the models adopted for each network component are provided in the following.

The linearized \ac{sg} model includes the linearization of the Sauer-Pai’s model \cite{Sauer1998}, of the steam turbine and governor \cite{mathworks2024steam,governor} and of the excitation system with the IEEE type 1 \ac{avr} \cite{mathworks2024excitation,avrIEEE}. It also includes the linearization of the rotation between the local and the global dq reference frame.

The linearized \ac{ibr} model is computed based on the description provided in Section~\ref{ssec:cig}. As in the case of the \ac{sg}, the rotation between local and global $dq$ reference frames is included in the linearization.

Transformers are modeled using the conventional equivalent T-circuit, which includes series RL branches for the primary and secondary windings, and a shunt branch consisting of the parallel of iron loss resistance and magnetizing inductance.

Loads are modeled by a constant RL series. Lines can be modeled using the conventional $\pi$-equivalent circuit by connecting an RL branch and two shunt capacitors. Clearly, no linearization is required for loads, lines and shunt capacitors.  

\begin{figure}[!ht]
    \centering
    \includegraphics[width=1\linewidth]{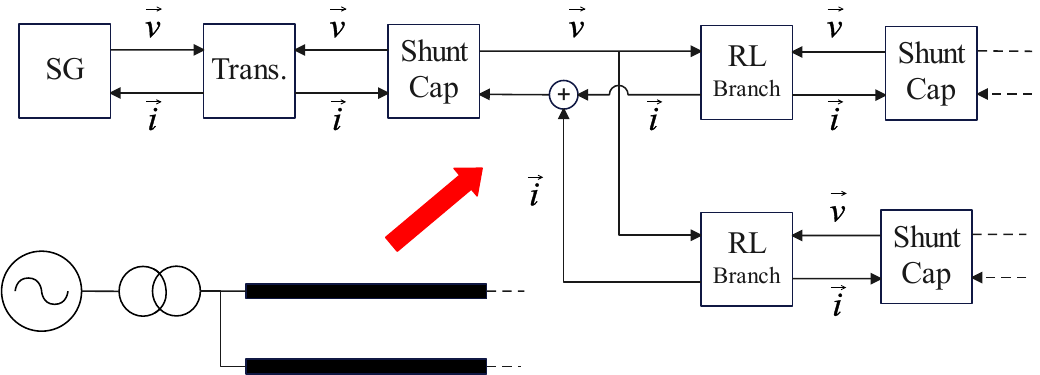}
    \caption{Linearization method, example of subsystems connection.}
    \label{fig:connection_example}
\end{figure}

To explain the final connection of the linear subsystems, we use the example reported in Fig.~\ref{fig:connection_example}. Here, we can observe how the small section of a grid in the bottom is represented by connecting the different subsystems. For example, the \ac{sg} subsystem receives the current from the transformer one and returns back the voltage; the transformer subsystem receives the voltage from the \ac{sg} and from the shunt capacitor (which is a part of the $\pi$-lines) and returns the currents; and so on. In case of multiple branches, as between the two lines and the transformer in the example case, the current connections must be handled by applying the Kirchhoff law. 

In mathematical terms, the interconnection of the subsystems is realized by a proper algebraic combination of their modeling matrices $A_{i},B_{i},C_i,D_i$ (where $i$ indexes the subsystems). The final result is the full power system linear model. In particular, the dynamical state matrix $A_{ps}$ can be used for the eigenvalue analysis.

This method was implemented on the Matlab platform and the interconnection of the subsystems was realized using the function \texttt{connect}.

To linearize, the initial values of the state variables of the nonlinear subsystems (i.e., \ac{sg} and \ac{ibr}) are required. We compute them starting from the result of a load flow executed by the open source tool \texttt{Pandapower} \cite{pandapower}.      

\section{Parameters stability region identification}
\label{ssec:asm}
Our main objective is to study the sensitivity of the power system stability with respect to the \ac{ibr} control parameters. For this purpose, we propose an \ac{asm} able to estimate the stability boundary of the power system in a given parameter space. 

Let $\rho \in \mathbb{R}^{N_{\rho}}$ be the vector collecting the considered parameters. We define the parameter space by intervals $[\rho_j^{min},\rho_j^{max}]$, $j=1,2,\ldots,N_{\rho}$. Then we define the function \texttt{isPSstable}($\rho$) that labels as stable ($s$=1) or unstable ($s$=0) the given parameter vector $\rho$ if the power system is stable in all the considered network scenarios. This means that one execution of \texttt{isPSstable}($\rho$) consists of the computation of the linearized model, as described in Section~\ref{ssec:linearization}, given the value of parameters in $\rho$ for all the considered scenarios. In our case study, this means that the function computes four different dynamical state matrices $A_{ps}$ and stability is stated if all of these matrices have not unstable eigenvalues (with nonnegative real part).

Being provided with the function \texttt{isPSstable}($\rho$) the parameter stability region can be ideally identified by scanning all the parameter space with a given discrete resolution. However, the computational burden of function \texttt{isPSstable}($\rho$) is significant, since it requires, at least, the multiple computation of the eigenvalues of a large matrix. 

To face this issue, we propose the \ac{asm} that proceeds as follows:

\begin{enumerate}
    \item  An initial random sampling set $G_{init}$ with uniform distribution and cardinality $N_{init}$ of the parameter space is generated. 
    \item Each sample $\rho_i$, $i=1,2,\ldots,N_{init}$ in  $G_{init}$ is labelled as stable ($s_i$=1) or unstable ($s_i$=0) using by function \texttt{isPSstable}($\rho_i$).
    \item The labeled samples are used to train a probabilistic classifier—specifically, a \ac{svm} with posterior calibration \cite{Cristianini2000} which estimates the probability of stability throughout the domain. This means that the probability of the power system stability can be predicted given a value $\rho$ of the parameters set through the trained classifier: $P_r^{SVM,0}(\rho)$.
    \item To refine stability boundaries, we then generate a large set of candidate points with cardinality $N_r>>N_{init}$ in the parameter space and predict their associated probability of stability through the trained \ac{svm}. Then, we select the $N_a$ candidate points closer to a predefined probability threshold $P_r^{th}$. These points are added to $G_{init}$  obtaining an augmented training set $G_{ai}$  (with cardinality $N_a+N_{init}$) and the \ac{svm} is re-trained, finally obtaining $P_r^{SVM}(\rho)$.
\end{enumerate}


It is clear that the use of this method, which yields probabilistic results, represents a trade-off between the accuracy of a deterministic analysis and the computational burden that the latter would entail.

The final trained classifier $P_r^{SVM}(\rho)$ can be used to identify the optimal parameter setting $\rho^* = \arg\max_{\rho} P_r^{SVM}(\rho)$, where optimality refers to the maximum estimated probability of stability.

\section{Results and discussion}
\label{sec:results}
The \ac{asm} is applied to the Cigr\'{e} case study described in Section~\ref{sec:Cigre_network} considering three different versions of the \ac{ibr} connected to bus 10: with an ideal dc voltage source at the dc port (version~1), with dc capacitor and $v_\mr{dc}$ control (version~2), and with dc capacitor and with $v_\mr{dc}^2$ control (version~3). The analysis is focused on the values \ac{pi} control gains of the \ac{pll} $(k_p^{pll},k_i^{pll})$, of the current control $(k_p^{cc},k_i^{cc})$ (assuming the same values for $i_d$ and $i_q$), and of the dc voltage control $(k_p^{(2)dc},k_i^{(2)dc})$.
The remaining parameters are defined as in Table~\ref{tab:ibr_parameters}, which also reports intervals for the control parameters used in the \ac{asm} (i.e., $[\rho_j^{min},\rho_j^{max}]$).
All values are defined with respect to the following nominal base values: 1500 \texttt{V} dc, 350 \texttt{MVA}, 660 \texttt{V}, 50~\texttt{Hz}.

\begin{table}[!t]
\begin{center}
\caption{IBR parameters\\Base values:  350~ \texttt{MVA} / 660 \texttt{V} / 50 \texttt{Hz},   1500 \texttt{V} dc.} 
\label{tab:ibr_parameters}
\begin{tabular}{lcc c}
\hline
\textbf{Description}                 & \textbf{Symbol} & \textbf{Value} & Unit \\ \hline
Filter resistance                    & $R$             &   0.05 & \texttt{pu}        \\
Filter inductance                    & $L$             &   0.15 & \texttt{pu}       \\
Filter capacitance                   & $C_f$           &   0.05 & \texttt{pu}       \\
Filter shunt resistance              & $R_f$           &   0.0016 & \texttt{pu}      \\ 
dc port capacitance                  & $C$             &   350 & \texttt{mF}           \\ \hline
PLL control proportional gain        & $k_p^{pll}$     &  [0 \ 12] & \texttt{pu}        \\
PLL control integral gain            & $k_i^{pll}$     &  [0 \ 860] & \texttt{pu}       \\
Current control proportional gain    & $k_p^{i}$       &  [0 \ 8] & \texttt{pu}      \\
Current control integral gain        & $k_i^{i}$       &  [0 \ 800] & \texttt{pu}      \\
$v_\mr{dc}$ control proportional gain   & $k_p^{dc}$      &  [0 \ 6] & \texttt{pu}          \\
$v_\mr{dc}$ control integral gain       & $k_i^{dc}$      &  [0 \ 1500] & \texttt{pu}      \\
$v_\mr{dc}^2$ control proportional gain & $k_p^{2dc}$     &  [0 \ 6.5] & \texttt{pu}               \\
$v_\mr{dc}^2$ control integral gain     & $k_i^{2dc}$     &  [0 \ 490] & \texttt{pu}             \\ \hline
\end{tabular}
\end{center}
\end{table}

The analysis is carried out using the \ac{asm} described in Section~\ref{ssec:asm}, with $P_r^{th}=0.8$, $N_{init}=100$, $N_a=250$.

We start from the \ac{ibr} version~1 applying the \ac{asm} over the 4D parameter space considering the \ac{pll} and the current control gains, $(k_p^{pll},k_i^{pll},k_p^i,k_i^i)$. The resulting trained classifier $P_r^{VSM}(\rho)$ is maximized to define a base case tuning of these parameters: $k_p^{pll,*}=0.77$~ \texttt{pu}, $k_i^{pll,*}=376$~ \texttt{pu}, $k_p^{i,*}=0.64$~ \texttt{pu}, $k_i^{i,*}=48$~ \texttt{pu}.

We remark that this tuning defines the control parameters values that should guarantee the maximal probability of stability over the four Cigr\'{e} grid scenarios, defined in Section~\ref{sec:Cigre_network}. Optimal tuning is not the main objective of this paper, since it should consider other features such as stability margin or damping. However, these values are necessary to show the result of our analysis, as it will appear clearer in the following. Indeed, visualizing the parameters' stability regions in a 4D space is not feasible. Therefore, we apply the \ac{asm} to the \ac{ibr} version~1 by considering a 2D parameter space at a time: first for the pair of \ac{pll} control gains and then for the pair of current control gains. In each case, the other pair is kept fixed at the optimal values previously computed.

\begin{figure}[!t]
    \centering
    \includegraphics[width=1\linewidth]{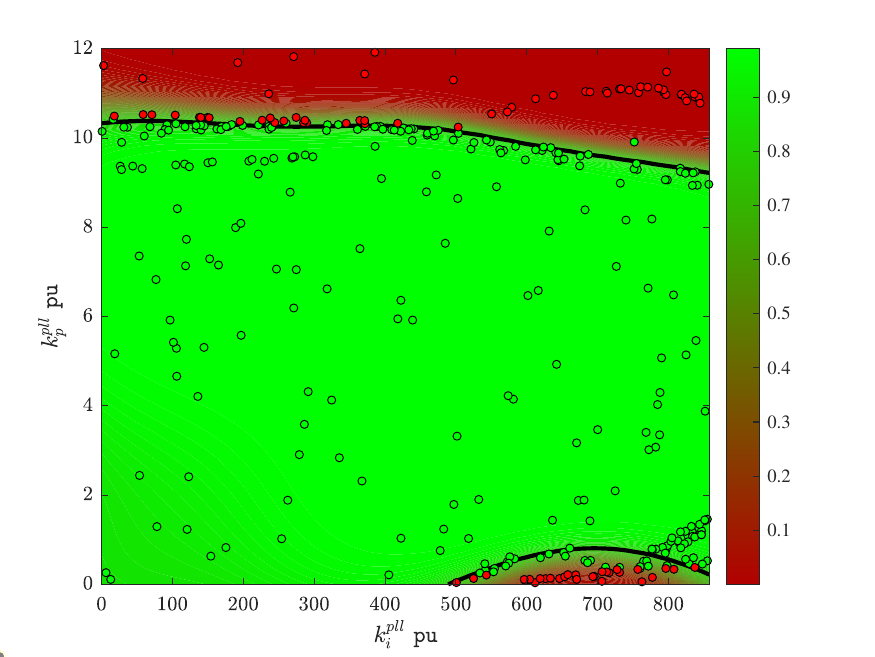}
    \caption{\ac{asm}-based analysis over the \ac{pll} control gains space.}
    \label{fig:ver0_pll_grid}
\end{figure}

Figure \ref{fig:ver0_pll_grid} shows the result of the analysis for the \ac{pll}  control gains. It shows the estimated probability of stability in the considered parameter space with full green indicating probability equal to one and full red indicating zero probability. The circles are the actual sampled values: if they are green, it means that the system was found to be stable for all the four Cigr\'{e} scenarios (\texttt{isPSstable}($\rho_i$) = 1); if they are red, it means that in at least one of the four scenarios the system resulted unstable (\texttt{isPSstable}($\rho_i$)=0). We can observe how the \ac{asm} increases the number of sampling around the boundary (where the estimate probability is $P_r^{th}=0.8$) in order to compute it with more accuracy. 

In Fig.~\ref{fig:ver0_pll_grid}, we can observe that the system is unstable when the proportional gain $k_p^{pll}$ is under a given threshold that increases with the value of the integral gain $k_i^{pll}$; whereas, if $k_p^{pll}$ is higher than a given value, almost independently of $k_i^{pll}$ the system is unstable. 

\begin{figure}[!t]
    \centering
    \includegraphics[width=1\linewidth]{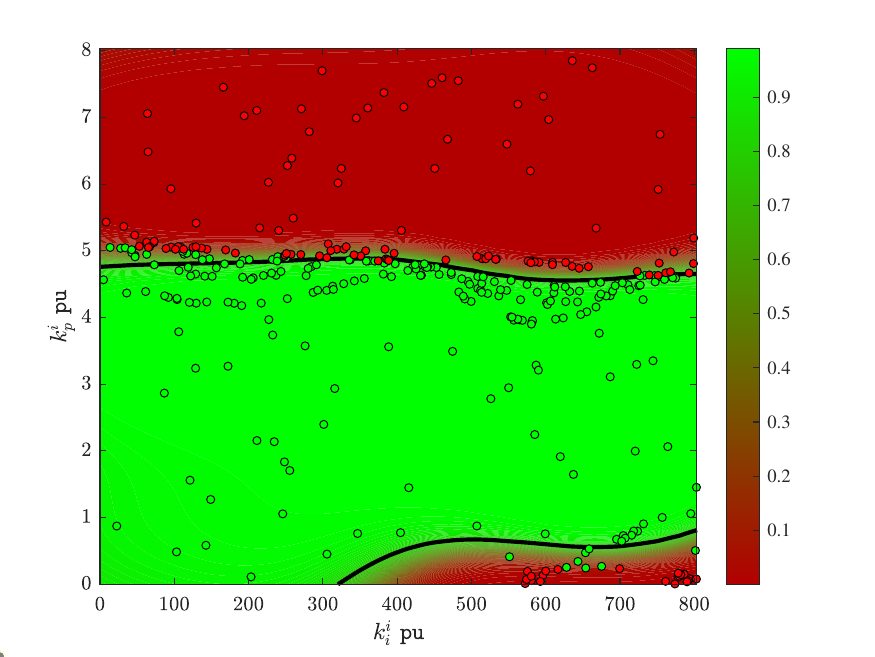}
    \caption{\ac{asm}-based analysis over the current control gains space.}
    \label{fig:ver0_cc_grid}
\end{figure}

Figure \ref{fig:ver0_cc_grid} shows the result of the analysis for the current control gains. In the figure, we can observe that the system is unstable if the proportional gain $k_p^{pll}$is higher than a given threshold, almost independently of the value of the integral gain $k_p^{pll}$; whereas if $k_p^{pll}$ is higher than 300  \texttt{pu}, $k_p^{pll}$ must be higher than a value that changes with $k_i^{pll}$.

We now continue the analysis by comparing the results obtained using the full linearized grid model with the ones get by representing the grid with the Thévenin equivalent. More specifically, we computed the Thévenin impedance and the Thévenin voltage for the four Cigr\'{e} grid scenarios starting from the system admittance matrix firstly including the loads and the shunt capacitors and then without including them. Table~\ref{tab:Thévenin_equivalents} reports the computed impedances in terms of short-circuit power $S_{sc}$ and X/R ratio and the Thévenin voltages in  \texttt{pu}. The table also reports the \ac{scr} defined with respect to the \ac{ibr} nominal power. We can observe that including the loads decreases the X/R ratio and increases $S_{sc}$ and the SCR. Conventionally, this means that the represented grid is stronger when loads are included in the computation.

\begin{table}[!t]
\caption{Thévenin equivalent parameters.}
\label{tab:Thévenin_equivalents}
\begin{center}
\begin{tabular}{llcccc}
\hline
Scenario & Loads Incl. & $S_{sc}$ \texttt{MVA} & SCR & X/R & $V_{th}$ \texttt{pu} \\ \hline
1 & Yes & 822 & 2.34 & 2.85 & 0.8201 \\
2 & Yes & 758 & 2.16 & 2.51 & 0.8259 \\
3 & Yes & 816 & 2.33 & 2.74 & 0.7984 \\
4 & Yes & 819 & 2.34 & 2.80 & 0.8068 \\ \hline
1 & No  & 692 & 1.98 & 7.26 & 1.0596 \\
2 & No  & 617 & 1.76 & 7.52 & 0.9348 \\
3 & No  & 676 & 1.93 & 7.39 & 1.0855 \\
4 & No  & 685 & 1.95 & 7.35 & 1.0788 \\ \hline
\end{tabular}
\end{center}
\end{table}

\begin{figure}[!t]
    \centering
    \begin{minipage}[b]{0.49\columnwidth}
        \centering
        \includegraphics[width=\columnwidth]{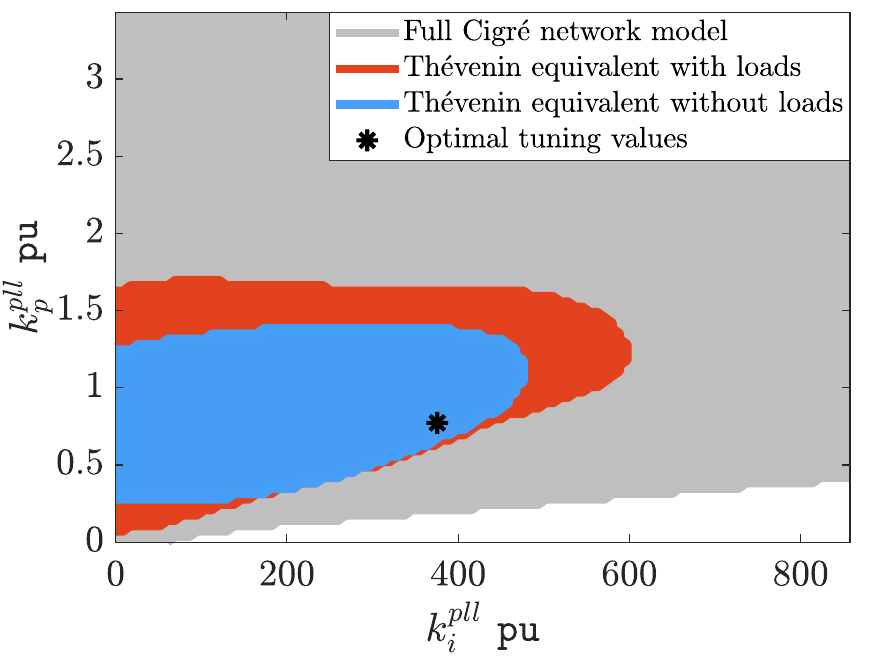}
        {\footnotesize \ac{ibr} Vers. 1: \ac{pll} control gains. \vspace{10pt}}
    \end{minipage}
    \hfill
    \begin{minipage}[b]{0.49\columnwidth}
        \centering
        \includegraphics[width=\columnwidth]{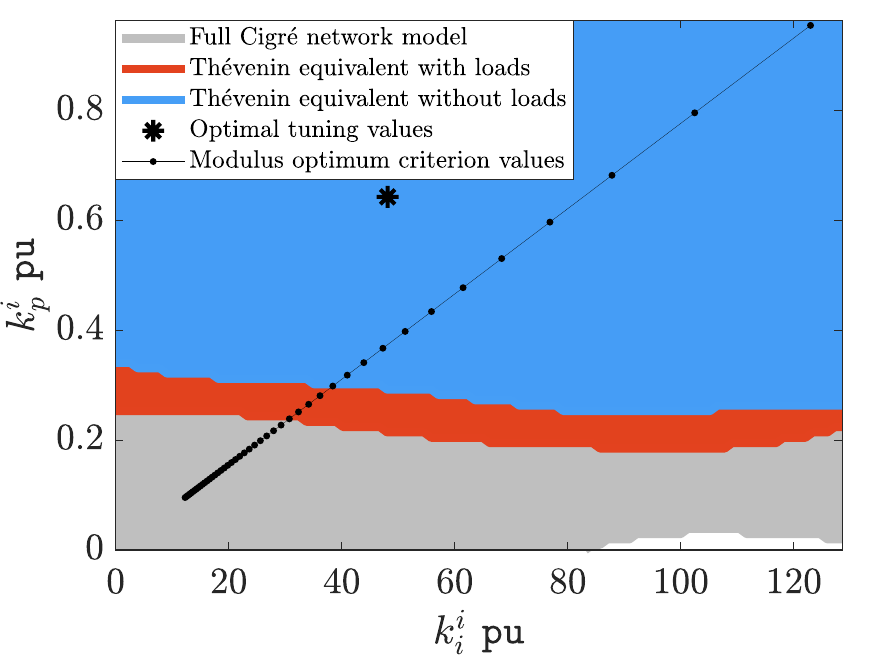}
        {\footnotesize \ac{ibr} Vers. 1: current control gains.    \vspace{10pt}}
    \end{minipage}
    \begin{minipage}[b]{0.49\columnwidth}
        \centering
        \includegraphics[width=\columnwidth]{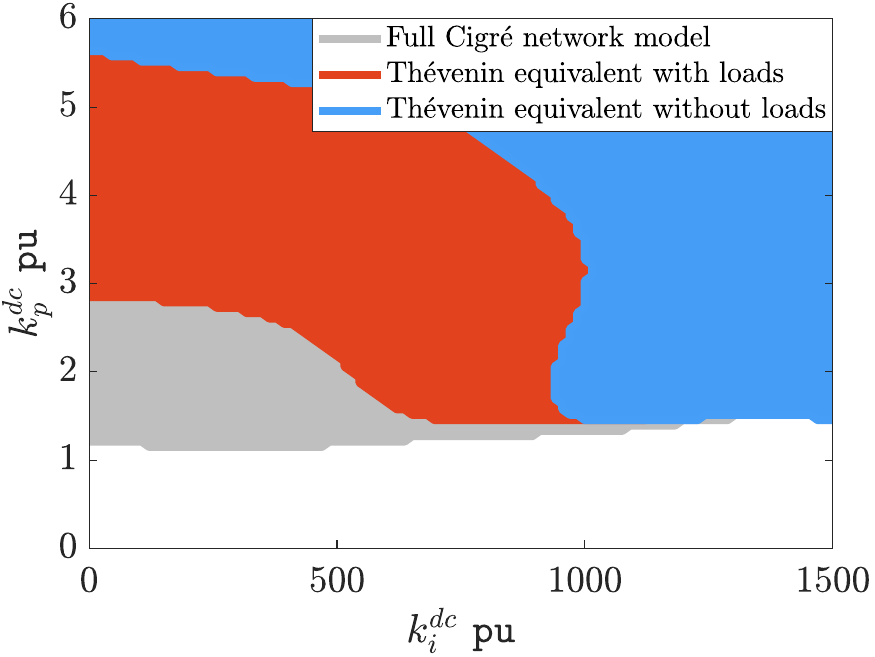}
        {\footnotesize \ac{ibr} Vers. 2: $v_\mr{dc}$ control gains.}
    \end{minipage}
    \hfill
    \begin{minipage}[b]{0.49\columnwidth}
        \centering
        \includegraphics[width=\columnwidth]{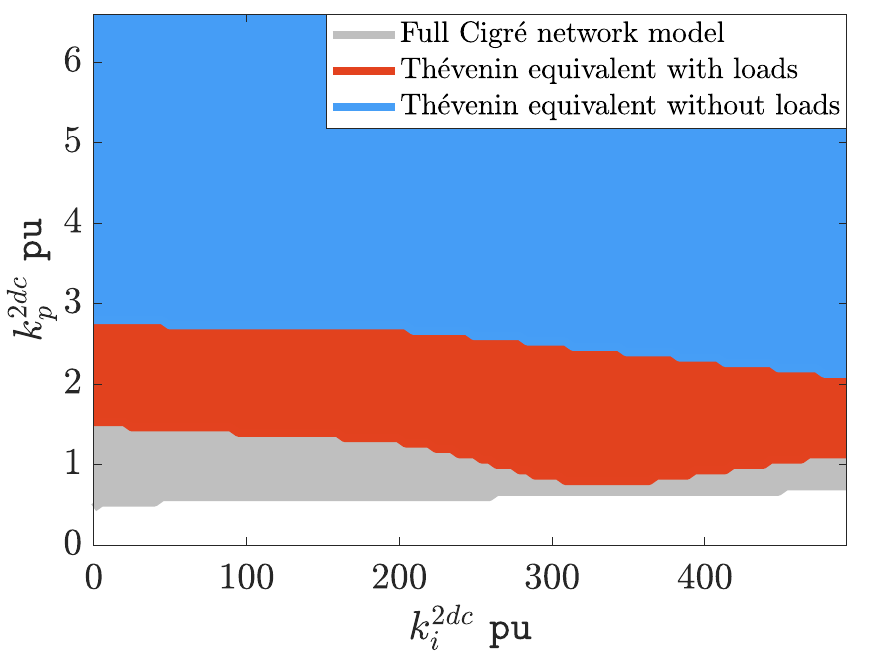}
        {\footnotesize \ac{ibr} Vers. 3: $v_\mr{dc}^2$ control gains.}
    \end{minipage}
    \caption{Parameters' stability regions with full grid model and equivalent Thévenin models.}
    \label{fig:stability_boundaries}
\end{figure}

Figure \ref{fig:stability_boundaries} shows the parameters' stability regions computed based on the three grid representations with: the \ac{ibr} version~1 over the \ac{pll} control gains, with the \ac{ibr} version~1 over the current control gains, with the \ac{ibr} version~2 over the $V_\mr{dc}$  control gains, and with the \ac{ibr} version~3 over the $V_\mr{dc}^2$ control. In the last two cases, the values of the \ac{pll} and current control gains are fixed to the mentioned optimal values.

For the analysis of \ac{ibr} version 1 with respect to the \ac{pll} control gains (top left in Fig.~\ref{fig:stability_boundaries}), we observe that the stability region computed using the Thévenin equivalent without considering the loads is contained within the one obtained when loads are included. Both regions are contained within the region computed using the full grid model. In the figure, we also report the optimal value computed using the \ac{svm} classifier based on the full linearized model, which results to be stable also when the grid is modeled by Thévenin equivalents.

It is worth noting how both Thévenin representations yield stability regions significantly different and smaller than the one obtained with the full grid model. This is not surprising, since the Thévenin representation is a strong simplification of the actual system model. Anyway, the fact that the Thévenin regions are contained within that of the full model indicates that the stability analysis based on the Thévenin equivalent is consistent and conservative, at least for this specific case and for the considered parameter range

A similar conclusion holds true for the case of  \ac{ibr} version~1 with respect to the current control gains (top right in Fig.~\ref{fig:stability_boundaries}). Here, we focus on the portion of the parameter space covered by the values that can be computed based on the tuning method referred to as the modulus optimum criterion \cite{yazdani2010}, which selects the PI gains as $k_p^i=L/\tau$ and $k_i^i=R/\tau$ with $\tau \in [0.5,5]$~\texttt{ms} controlling the regulator's response time. In this portion, according to the full model, stability is almost always guaranteed (excepting for a few values of $k_i^i$ between 90~\texttt{pu} to 130~\texttt{pu}, for which the proportional gain must be higher than zero). Differently, using the Thévenin equivalents, the conclusion is that the proportional gain should be higher than a given threshold, which is only slightly dependent on the value of the integral gain. This threshold result to be generally higher, and thus more conservative, with the Thévenin representation without loads.

Finally, in the two bottom graphs in Fig.~\ref{fig:stability_boundaries} we can analyze the results obtained over the parameter spaces of the dc voltage control gains, in the cases of $v_\mr{dc}$ and $v_\mr{dc}^2$ control. As for the \ac{ibr} version~1, we obtain that the regions computed based on the Thévenin equivalents are contained in the one computed using the full grid model. For the $v_\mr{dc}$ control gains (\ac{ibr} version~2), the full grid model suggests that stability is ensured if the proportional gain exceeds a certain threshold, with little dependence on the integral gain. In contrast, the Thévenin equivalent models reveal a stronger interdependence between the proportional and integral gains to guarantee stability. 

In the case of the $v_\mr{dc}^2$ control (\ac{ibr} version~3), although a stronger interdependence between the proportional and integral gains is observed, the shape of the stability regions obtained using the Thévenin equivalents closely resembles that derived from the full model. All indicates the existence of a minimum required value for the proportional gain, which results to be higher when Thévenin equivalents are adopted.

\section{Conclusions}
\label{sec:Conclusions}
This paper has presented a detailed small-signal stability analysis of a modified version of the Cigré HV European network under the penetration of grid-following IBRs. Stability is assessed considering linearized versions of highly detailed models of power equipment and associated controls, using an ASM to make the approach practical in an environment with limited computational resources. The methodology facilitates the selection of filter elements or control parameter tuning for any new device entering an existing power system.        

The focus of the paper is on the influence that grid-following IBRs' control parameters have on the overall grid stability, under various levels of mechanical inertia. The study also provides insight on the accuracy of simplified models based on Thévenin equivalents, highlighting the importance of knowing the true value of Thévenin parameters in order to make accurate predictions about grid level stability.

The authors are currently expanding the approach to include grid-forming IBRs, different IBR control schemes, and more complex realizations of IBRs' penetration. Results will be published as they become available.

\bibliographystyle{IEEEtran}

\end{document}